# Performances Evaluation of Enhanced Basic Time Space Priority combined with an AQM

Said EL KAFHALI,  Mohamed HANINI,  Abdelali EL BOUCHTI,  Abdelkrim HAQIQ

Computer, Networks, Mobility and Modeling laboratory
Department of Mathematics and Computer
FST, Hassan 1st University, Settat, Morocco
e-NGN Research group, Africa and Middle East

*Abstract*— **Active Queue Management( AQM) is an efficient tool in the network to avoid saturation of the queue by warning the sender that the queue is almost full to reduce its speed before the queue is full. The buffer management schemes focus on space management, in the other hand scheduling priorities (focusing on time management) attempt to guarantee acceptable delay boundaries to applications for which it is important that delay is bounded.**

**Combined mechanisms (time and space management) are possible and enable networks to improve the perceived quality for multimedia traffic at the end users.**

**The key idea in this paper is to study the performance of a mechanism combining an AQM with a time-space priority scheme applied to multimedia flows transmitted to an end user in HSDPA network. The studied queue is shared by Real Time and Non Real Time packets.**

**We propose a mathematical model using Poisson and MMPP processes to model the arrival of packets in the system. The performance parameters are analytically deducted for the Combined EB-TSP and compared to the case of Simple EB-TSP.**

**Numerical results obtained show the positive impact of the AQM added to the EB-TSP on the performance parameters of NRT packets compared to the Simple EB-TSP.**

## I. INTRODUCTION

To avoid congestion in high-speed networks, due to increased traffic which transits among them, we use buffers (Queues) in routers to handle the excess of traffic when the debit exceeds the transmission capacity. But the limited space of these buffers, cause the loss of packets of information over time. Management mechanisms queues have great utilities to avoid buffers congestion. These mechanisms differ in the method of selection of discarded packets. We distinguish two categories of mechanisms: passives mechanisms (PQM: Passive Queue Management) that detects congestion only after a packet has been dropped at the gateway and actives mechanisms (AQM: Active Queue Management) that takes a preventive approach by removing packets before the saturation of the buffer, and this with a probability depending on the size of the queue. This allows avoid saturation of the queue warning the sender that the queue is almost full to reduce its speed and drops packets before the queue is full. Several AQM has been proposed in the literature, Floyd and Jacobson proposed the RED algorithm (Random Early Detection) [9].

The RED calculates the average queue size, using a low-pass filter with an exponential weighted moving average. The average queue size is compared to two thresholds, a minimum threshold and a maximum threshold. When the average queue size is less than the minimum threshold, no packets are marked. When the average queue size is greater than the maximum threshold, every arriving packet is marked. If marked packets are in fact dropped, or if all source nodes are cooperative, this ensures that the average queue size does not significantly exceed the maximum threshold [7], [9].

The buffer management schemes focus on space management. In the other hand scheduling priorities referred as time priority schemes attempt to guarantee acceptable delay boundaries to real time (RT) applications (voice or video) for which it is important that delay is bounded.

Combined mechanisms (time and space management) are possible and enable networks to improve the perceived quality for multimedia traffic at the end users.

Work in [3] present a queuing model for multimedia traffic over HSDPA channel using a combined time priority and space priority (TS priority) with threshold to control QoS measures of the both RT and NRT packets.

The basic idea is that, in the buffer, RT packets are given transmission priority (time priority), but the number accepted of this kind of packets is limited. Thus, this scheme aims to provide both delay and loss differentiation.

Authors in [16] show, via simulation (using OPNET), that the TSP scheme achieves better QoS measures for both RT and NRT packets compared to FCFS (First Come First Serve) queuing.

To model the TSP priority scheme, mathematical tools have been used in ([4], [5], [14]) and QoS measures have been analytically deducted.

When the TSP scheme is applied to a buffer in Node B (in HSDPA technology) arriving RT packets will be queued in front of the NRT packets to receive priority transmission on the shared channel. A NRT packet will be only transmitted when no RT packets are present in the buffer, this may the RT QoS delay requirements would not be compromised [2].

In order to fulfil the QoS of the loss sensitive NRT packets, the number of admitted RT packets, is limited to $R$ , to devote more space to the NRT flow in the buffer.

Authors in [11] present and study an enhancement of the TS priority (EB-TSP) to overcome a drawback of the scheme presented in [3]: bad QoS management for RT packets, and bad management for buffer space.

In order to show the importance of the AQM mechanisms to improve the Quality of Service (QoS) in the HSDPA networks, we propose in this work to combine the EB-TSP scheme with a mechanism to control the arrival rate of NRT packets in the buffer.

Hence, in this paper two mechanisms are compared. In the first mechanism, called Simple EB-TSP (S-EB-TSP), the both type of packets are not controlled. But in the second mechanism, called Combined EB-TSP (C-EB-TSP), an AQM is used to control the NRT packets. The RT arrivals are modeled by an MMPP process and the NRT arrivals by a Poisson process.

Our main objective is to present and compare two queue management mechanisms with a time and space priority scheme for an end user in HSDPA network. Those mechanisms are used to manage access packets in the queue giving priority to the Real Time (RT) packets and avoiding the Non Real Time (NRT) packet loss.

A queuing analytical model is presented to evaluate the performance of both mechanisms. A discrete time Markov chain is formulated by considering Markov Modulated Poisson Process (MMPP) as the traffic source of RT packets. The advantages of using MMPP are two-fold: first, MMPP is able to capture burstiness in the traffic arrival rate which is a common characteristic for multimedia and real-time traffic sources as well as Internet traffic [13]. Second, it is possible to obtain MMPP parameters analytically for multiplexed traffic sources so that the queueing performances for multiple flows can be analyzed.

A dynamic access control (AQM) for the NRT packets in the channel is added in the second mechanism. This mechanism should determine dynamically the number of the NRT packets accepted in the queue instead of to fix it at the beginning.

The rest of the paper is as follows. Section II gives an idea about MMPP process and describes mathematically the two mechanisms. In Section III, we present the performance parameters of these mechanisms. Numerical results are contained in Section IV and section V concludes the paper.

## II. FORMULATION OF THE ANALYTICAL MODEL

### A. The Markov Modulated Poisson Process

The Markov Modulated Poisson Process (MMPP) is a term introduced by Neuts [15] for a special class of versatile point processes whose Poisson arrivals are modulated by a Markov process. The model is a doubly stochastic Poisson process [8], whose rate varies according to a Markov process; it can be used to model time-varying arrival rates and important correlations between inter-arrival times. Despite these abilities, the MMPPs are still tractable by analytical methods.

The current arrival rate $\lambda_i$ , $1 \leq i \leq S$ of an MMPP is defined by the current state $i$ of an underlying Continuous Time Markov Chain (CTMC) with $S$ states. The counting process of an MMPP is given by two processes $\{(J(t), N(t) : t \in T\}$ , where $N(t)$ is the number of arrivals within certain is time interval $[0, t)$ , $t \in T$ and $1 \leq J(t) \leq S$ is the state of the underlying CTMC. Also, the MMPP parameters can be represented by the transition probability matrix of the modulating Markov chain $\Psi$ and the Poisson arrival rate matrix $A$ as follows:

$$\Psi = \begin{bmatrix} \psi_{11} & \cdots & \psi_{1S} \\ \vdots & \ddots & \vdots \\ \psi_{S1} & \cdots & \psi_{SS} \end{bmatrix}, A = \begin{bmatrix} \lambda_1 & & \\ & \ddots & \\ & & \lambda_s \end{bmatrix} \qquad (1)$$

The rates of the transitions between the states of the CTMC are given by the non-diagonal elements of $\Psi$ .

The steady-state probabilities of the underlying Markov chain $\pi_\psi$ are given by:

$$\pi_\psi = \pi_\psi . \Psi \text{ and } \pi_\psi . 1 = 1 \qquad (2)$$

where $1$ is a column matrix of ones.

The mean steady state arrival rate generated by the MMPP is:

$$\overline{\lambda} = \pi_\psi . \lambda^T \qquad (3)$$

where $\lambda^T$ is the transpose of the row vector $\lambda = (\lambda_1, \ldots, \lambda_S)$ .

### B. Mechanisms description

For the two mechanisms studied in this paper, we model the HSDPA link by a single queue of finite capacity $N$, $N>0$. The arriving flow in the queue is heterogeneous and composed by the RT and NRT packets.

The arrivals process of the RT packets are modeled by a 2-state MMPP characterized by the arrival Poisson rates $\lambda_1$ and $\lambda_2$ and the transition rates between them. We denote $\sigma_1$ and $\sigma_2$ the transition rate from $\lambda_1$ to $\lambda_2$ and transition rate from $\lambda_2$ to $\lambda_1$ respectively.

The average arrival rate for the RT packets modeled by a 2-state MMPP is calculated by:

$$\lambda_{average} = \frac{\sigma_1.\lambda_2 + \sigma_2.\lambda_1}{\sigma_1 + \sigma_2} \qquad (4)$$

The arrivals process of the NRT packets are modeled by Poisson process with rate $\lambda$ .

As described in [11] the access to the buffer is determined by the following policy:
When an RT packet arrives at the buffer, either it is full or there is free space. In the first case, if the number of RT packets is less than, then an NRT packet will be rejected and the arriving RT packet will enter in the buffer. Or else, the arriving RT packet will be rejected. In the second case, the arriving RT packet will enter in the buffer.
The same, when an NRT packet arrives at the buffer, either it is full or there is free space. In the first case, if the number of RT packets is less than $R$ , then the arriving NRT packet will be rejected. Or else, an RT packet will be rejected and the arriving NRT packet will enter in the buffer. In the second case, the arriving NRT packet will enter in the buffer.

In the queue, the server changes according to the type of packet that it treats, a server is reserved for the RT packets and another for the NRT packets; these two servers operate independently. Furthermore, we assume that the server is exponential with parameter $\mu$ (respectively $\mu_1$) for the RT packets (respectively for the NRT packets).

In the first mechanism (Figure 1), called S-EB-TSP, the NRT packets arrive according to a Poisson process and their number in the queue cannot exceed $N$ .

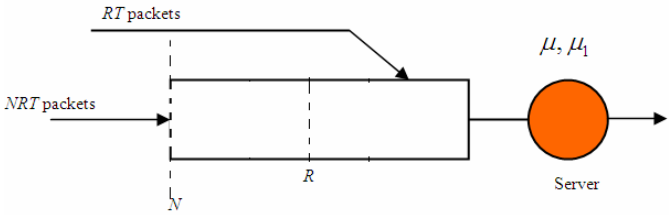

Figure 1: System with Simple EB-TSP

In the second mechanism (Figure 2), called C-EB-TSP, we add two other thresholds $L$ and $H$ ( $R < H < L$ such that $L = N - R$ ) in the queue in order to control the arrival rate of the NRT packets.
Let $k$ be the total number of packets in the queue at time $t$ .

- If $0 \le k < H$ , then the arrival rate of the NRT packets is $\lambda$ .

- If $H \le k < L$ then the arrival rate of NRT packets is reduced to $\dfrac{\lambda}{2}$

- If $k \ge L$ , then no NRT packets arrives in the queue. This can be considered as an implicit feedback from queue to the Node B.

This second mechanism enables to prevent either the congestion in the system or the loss of the NRT packets.

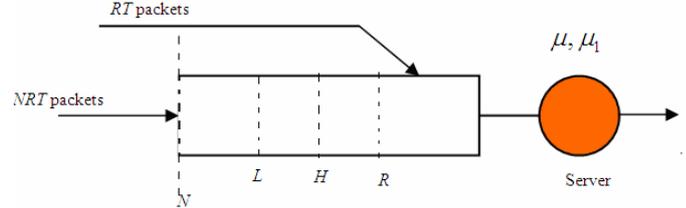

Figure 2: System with Combined EB-TSP

**Remark**: In the buffer, the RT packets are placed all the time in front of the NRT packets.

### C. Mathematical description

For the first mechanism, the state of the system is described at time $t (t \ge 0)$ by the stochastic process $X_t = \left( X, X_t^1, X_t^2 \right)$, where $X$ is the phase of the MMPP and $X_t^1$ (respectively $X_t^2$ ) is the number of the RT (respectively NRT) packets in the queue at time $t$ .
The state space of $X_t$ is

$$E_1 = \{1, 2\} \times \{0, ....., R\} \times \{0, ....., N\} \qquad (5)$$

For the second mechanism, the state of the system is described at time $t (t \ge 0)$ by the stochastic process $Y_t = \left( Y, Y_t^1, Y_t^2 \right)$, where $Y$ is the phase of the MMPP and $Y_t^1$ (respectively $Y_t^2$ ) is the number of the RT (respectively NRT) packets in the queue at time $t$ .
The state space of $Y_t$ is:

$$E_2 = \{1, 2\} \times \{0, ....., R\} \times \{0, ....., H\} \qquad (6)$$

### III. STATIONARY PROBABILITIES AND PERFORMANCE PARAMETERS

#### A. Stationary Pribabilities

For the both systems, the inter-arrival times are exponential. The service times are exponentials. And all these variables are mutually independent between them, then $X_t$ and $Y_t$ are Markov process with finite state spaces (because the exponential is without memory).

We also remark that the process $X_t$ and $Y_t$ are irreducible (all their states communicate between them). Thus, we deduct that $X_t$ and $Y_t$ are ergodic (i.e. the systems are stable).

Consequently, the stationary probabilities of $X_t$ and $Y_t$ exist and can be computed by solving the system of the balance equations (the average flow outgoing of each state is equal to the average flow go into state) in addition to the normalization equation (the sum of all state probabilities equal to 1).

Let $p_1(i, j, k)$ (respectively $p_2(i, j, k)$ ) denotes the stationary probability for the state $(i, j, k)$ where $(i, j, k) \in E_1$ (respectively $(i, j, k) \in E_2$ ).

#### B. Performance Parameters

In this section, we determine analytically, different performance parameters (loss probability of the RT packets, average numbers of the RT and NRT packets in the queue, and average delay for the RT and NRT packets) at the steady state. These performance parameters can be derived from the stationary state probabilities as follows:

#### B.1 System with Simple EB-TSP

##### a) Loss probability of the RT packets :

Using the ergodicity of the system, the loss probability of *RT* packets for system with Simple EB-TSP is given by:

$$P^S_{Loss-RT} = \frac{\sum_{i=1}^{2} \lambda_i \sum_{j=0}^{N} p_1(i, j, N-i) + \lambda_{NRT} \sum_{i=1}^{2} \sum_{j=R+1}^{N} p_1(i, j, N-i)}{\frac{\sigma_2 . \lambda_1 + \sigma_1 . \lambda_2}{\sigma_1 + \sigma_2}} \quad (7)$$

##### b) Loss probability of the NRT packets :

Using a same analysis, we can show that the loss probability of NRT packets is:

$$P^S_{Loss-NRT} = \frac{\sum_{i=1}^{2} \sum_{j=0}^{R-1} \lambda_i . p_1(i, j, N-i)}{\lambda} + \sum_{i=1}^{2} \sum_{j=0}^{R} p_1(i, j, N-i) \quad (8)$$

##### c) Average numbers of the RT and NRT packets in the queue:

There are obtained as follows:

$$N^S_{RT} = \sum_{i=1}^{2} \sum_{j=0}^{N} \sum_{k=0}^{N-i} j p_1(i, j, k) \quad (9)$$

$$N^S_{NRT} = \sum_{i=1}^{2} \sum_{k=0}^{N} \sum_{j=0}^{N-k} j p_1(i, k, j) \quad (10)$$

##### d) Average Packets Delay

It is defined as the number of packets waits in the queue since its arrival before it is transmitted. We use Little's law [14] to obtain respectively the average delays of RT and NRT packets in the system as follows:

$$D^S_{RT} = \frac{N^S_{RT}}{\lambda_{avg-RT} (1 - P^S_{loss-RT})} \quad (11)$$

$$D^S_{NRT} = \frac{N^S_{RT} + N^S_{NRT}}{\lambda (1 - P^S_{loss-NRT})} \quad (12)$$

Where $$\lambda_{avg-RT} = \frac{\sigma_2 . \lambda_1 + \sigma_1 . \lambda_2}{\sigma_1 + \sigma_2} \quad (13)$$

#### B.2 System with Combined EB-TSP

##### a) Loss probability of the RT packets :

The loss probability of RT packets is given by:

$$P^C_{Loss-RT} = \frac{\sum_{i=1}^{2} \lambda_i \sum_{j=0}^{N} p_2(i, j, N-i) + \frac{\lambda}{2} \sum_{i=1}^{2} \sum_{j=R+1}^{N} p_2(i, j, N-i)}{\frac{\sigma_2 . \lambda_1 + \sigma_1 . \lambda_2}{\sigma_1 + \sigma_2}} \quad (14)$$

##### b) Loss probability of the NRT packets :

The loss probability of NRT packets is given by:

$$P^C_{Loss-NRT} = \frac{\sum_{i=1}^{2} \sum_{j=0}^{R-1} \lambda_i . p_2(i, j, N-i)}{\lambda / 2} \quad (15)$$

*c) Average number of the RT packets in the queue:*
It is given by:

$$N_{RT}^C = \sum_{i=1}^{2} \sum_{j=0}^{N} \sum_{k=0}^{N-i} j p_2(i,j,k) \qquad (16)$$

*d) Average numbes of the NRT packets in the queue:*
It is given by:

$$N_{NRT}^C = \sum_{i=1}^{2} \sum_{k=0}^{N} \sum_{j=0}^{N-k} j p_2(i,k,j) \qquad (17)$$

*e) Average delays of the RT and NRT packets in the queue:*

$$D_{RT}^C = \frac{N_{RT}^C}{\lambda_{avg-RT}(1 - P_{loss-RT}^C)} \qquad (18)$$

$$D_{NRT}^C = \frac{N_{RT}^C + N_{NRT}^C}{\lambda_{NRT-eff}(1 - P_{loss-NRT}^C)} \qquad (19)$$

Where : $\lambda_{NRT-eff}$ is the effective arrival rate of NRT packets. It is computed by following formula:

$$\lambda_{NRT-eff} = \lambda . \sum_{i=1}^{2} \sum_{j=0}^{R} k p_2(i,j,k) + \frac{\lambda}{2} . \sum_{i=1}^{2} \sum_{j=0}^{R} \sum_{k=0}^{H-i-1} k p_2(i,j,k) \quad (20)$$

## IV.  NUMERICAL RESULTS

In [11], the authors have just calculated and evaluated the performance parameters for the mechanism called Simple EB-TSP. Here we present a comparison between the first and second mechanisms.

We remark that both mechanisms present similar performances for the RT packets. Whereas, the performances for the NRT packets vary from a mechanism to the other. Furthermore, to see the difference between the performance parameters of the NRT packets for both mechanisms, we study some simulations below.

For $N = 60$ , $H = 25$ , $L = 45$ , $R = 15$ , $\lambda = 20$ , $\lambda_1 = 8$ , $\lambda_2 = 5$ and $\mu_1 = 20$ , we remark that when the service rate $\mu$ of the RT packets increases, the average delay and the average number of the NRT packets are lower in the second mechanism than in the first mechanism (Figure 3) and when $\mu$ is lower the second mechanism is clearly more effective.

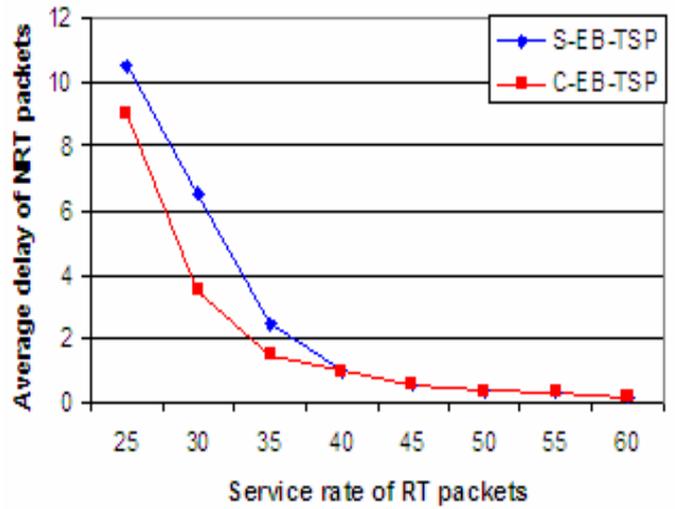

Figure 3: Average delays of NRT packets versus service rate of RT packets

For $N = 60$ , $H = 25$ , $L = 45$ , $R = 15$ , $\lambda = 20$ , $\lambda_1 = 8$ , $\lambda_2 = 5$ and $\mu = 30$ , the same behavior of the average delay of the NRT packets is shown in figure 4, which represents the variations of this performance parameter according to the service rate of the NRT packets.

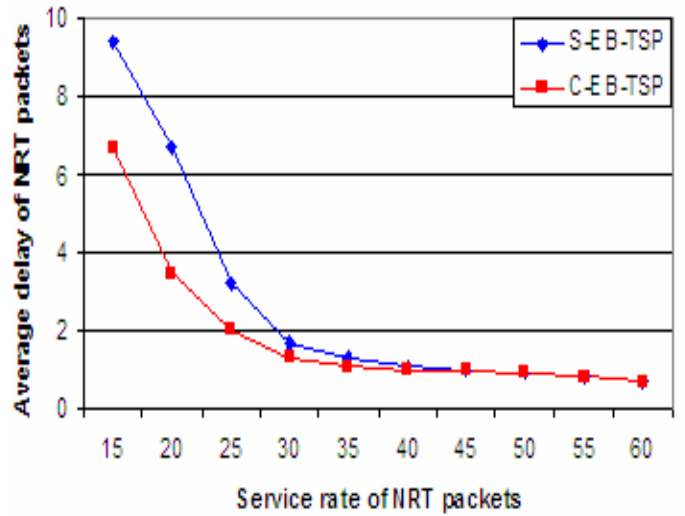

Figure 4:  Average delays of NRT packets versus service rate of NRT packets.

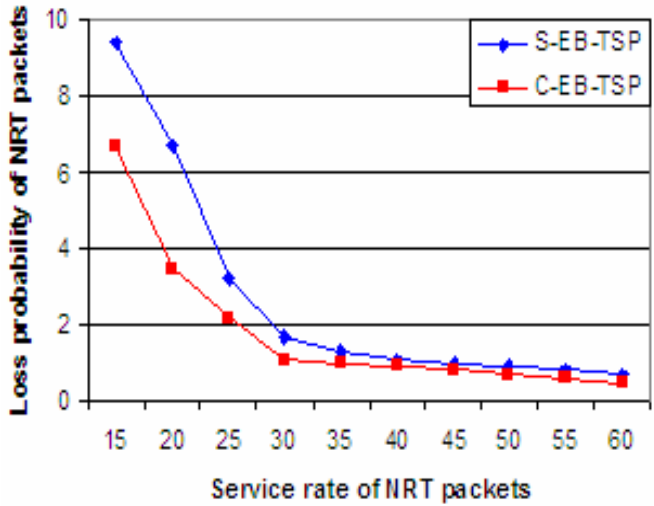

Figure 5: Loss probability of NRT packets versus service rate of NRT packets.

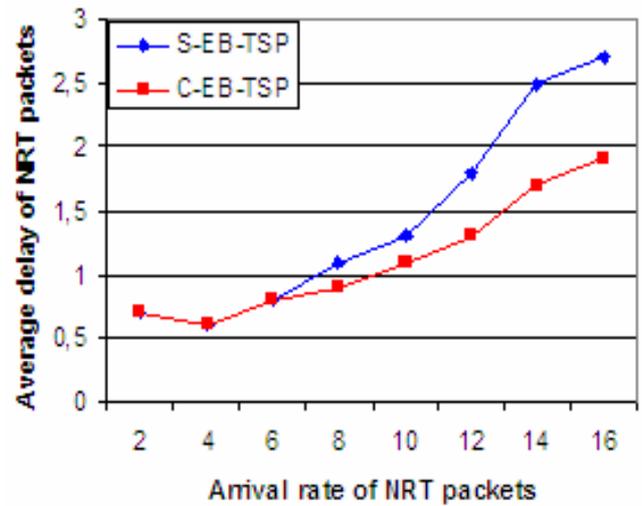

Figure 7: Average delay of NRT packets according to the arrival rate of NRT packets.

In Figure 5, we remark that the second mechanism where the EB-TSP scheme is combined with an AQM achieves a gain on the loss probability of NRT packets.

For $N = 60$ , $H = 25$ , $L = 45$ , $R = 15$ , $\lambda = 8$ $\mu = 30$ and $\mu_1 = 25$ , We remark that when the arrival rate of the RT packets increases, the average delay and the average number of the NRT packets are lower in the second mechanism. When the arrival rate off RT packets is higher the second mechanism enhances these parameters (Figure 6).

Figure 7 compares the behavior of the delay of NRT packets in the two mechanisms when the arrival rate of the NRT packets varies and shows that the second mechanism is more effective, especially when $\lambda$ is higher.

## V. CONCLUSION

The key idea in this paper is to study the performance of a mechanism combining an AQM with a time-space priority scheme applied to multimedia flows transmitted to an end user in HSDPA network. The studied queue is shared by Real Time and Non Real Time packets.

Mathematical tools are used in this study, we use Poisson and MMPP processes to model the arrival of packets in the system, and performance parameters are analytically deduced for the Combined EB-TSP and compared to the case of simple EB-TSP.

Numerical results obtained show that the performance parameters of *RT* are similar in the two mechanisms, where as the C-EB-TSP where the AQM is combined with the time-Space priority scheme achieves better performances for NRT packets compared to the Simple Eb-TSP.

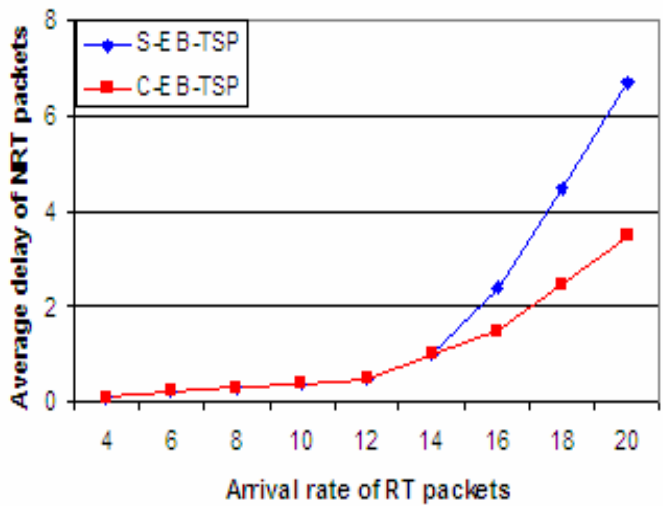

Figure 6: Average delays of NRT packets according to the arrival rate of RT packets.

## AUTHORS PROFILE

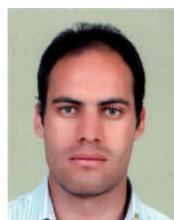

**Said EL KAFHALI** received the B.Sc. degree in Computer Sciences from the University of Sidi Mohamed Ben Abdellah, Faculty of Sciences Dhar El- Mahraz, Fez, Morocco, in 2005, and a M.Sc. degree in Mathematical and Computer engineering from the Hassan 1st University, Faculty of Sciences and Techniques (FSTS), Settat, Morocco, in 2009. He has been working as professor of Computer Sciences in high school since 2006, Settat, Morocco. Currently, he is working toward his Ph.D. at FSTS. His current research interests performance evaluation, analysis and simulation of Quality of Service in mobile networks.

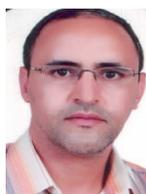

**Mohamed HANINI** is currently pursuing his PhD. Degree in the Department of Mathematics and Computer at Faculty of Sciences and Techniques (FSTS), Settat, Morocco. He is member of e-ngn research group. His main research areas are: Quality of Service in mobile networks, network performance evaluation.

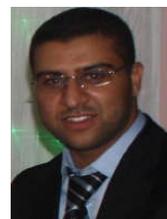

**Abdelali EL BOUCHTI** received the B.Sc. degree in Applied Mathematics from the University of Hassan 2nd, Faculty of Sciences Ain chock, Casablanca, Morocco, in 2007, and M.Sc. degree in Mathematical and Computer engineering from the Hassan 1st University, Faculty of Sciences and Techniques (FSTS), Settat, Morocco, in 2009. Currently, he is working toward his Ph.D. at FSTS. His current research interests include performance evaluation and control of telecommunication networks, stochastic control, networking games, reliability and performance assessment of computer and communication systems.

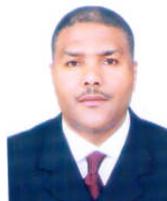

**Dr. Abdelkrim HAQIQ** has a High Study Degree (DES) and a PhD (Doctorat d'Etat) both in Applied Mathematics from the University of Mohamed V, Agdal, Faculty of Sciences, Rabat, Morocco. Since September 1995 he has been working as a Professor at the department of Mathematics and Computer at the faculty of Sciences and Techniques, Settat, Morocco. He is the director of Computer, Networks, Mobility and Modeling laboratory and a general secretary of e-NGN research group, Moroccan section. He was the chair of the second international conference on Next Generation Networks and Services, held in Marrakech, Morocco 8 – 10 July 2010.

Professor Haqiq' interests lie in the area of applied stochastic processes, stochastic control, queueing theory and their application for modeling/simulation and performance analysis of computer communication networks.

From January 98 to December 98 he had a Post-Doctoral Research appointment at the department of systems and computers engineering at Carleton University in Canada. He also has held visiting positions at the High National School of Telecommunications of Paris, the universities of Dijon and Versailles St-Quentin-en-Yvelines in France, the University of Ottawa in Canada and the FUCAM in Belgium.